\documentclass[iop,apj]{emulateapj}
\usepackage{amsmath}  
\usepackage{subfigure}

\epsscale{1.3}

\newcommand{\Aedg}    {\ensuremath{A_\textrm{edge}}}    
\newcommand{\aeff}    {\ensuremath{a_\textrm{eff}}}

\newcommand{\aN}      {\ensuremath{a_\textrm{N}}}
\newcommand{\app}     {\ensuremath{\sim}}
\newcommand{\avg}[1]  {\ensuremath{\overline{#1}}} 
\newcommand{\aZ}      {\ensuremath{a_0}}
\newcommand{\Cf}      {\ensuremath{C_\textrm{f}}}       
\newcommand{\dex}[1]  {\ensuremath{\times\textrm{10}^{#1}}}
\newcommand{\G}       {\ensuremath{ \textrm{G} }}
\newcommand{\Ha}        {\textrm{H}\ensuremath{ {\alpha}}}
\newcommand{\HI}      {\textrm{H{\sc i}}}
\newcommand{\Htwo}    {\textrm{H}\ensuremath{_{2}}\ }
\newcommand{\kms}     {\ensuremath{ \textrm{km~s}^{-1}}}
\newcommand{\kpc}     {\textrm{kpc}}
\newcommand{\LCDM}    {\ensuremath{\Lambda\textrm{CDM}}}

\newcommand{\Lsun}    {\ensuremath{ \textrm{L}_{\sun}~}}
\newcommand{\lt}      {\ensuremath{<}}
\newcommand{\Lvec}    {\ensuremath{\textbf{L}}}
\newcommand{\Lzvec}   {\ensuremath{\textbf{L}_{0}}}
\newcommand{\Mg}      {\ensuremath{M_\textrm{g}}}
\newcommand{\MHI}     {\ensuremath{ \textrm{M}_\HI}}
\newcommand{\ML}      {\ensuremath{\Upsilon_\star}}
\newcommand{\MLnir}   {\ensuremath{\Upsilon_\star^{[3.6]}}}
\newcommand{\MLI}     {\ensuremath{\Upsilon_\textrm{I}}}
\newcommand{\mm}      {\ensuremath{{\mu{m}}}}
\newcommand{\Ms}      {\ensuremath{M_{\star}}}
\newcommand{\Msun}    {\textrm{M}\ensuremath{_{\odot}}}
\newcommand{\Mt}      {\ensuremath{M_\textrm{t}}}
\newcommand{\muB}     {\ensuremath{\mu_\textrm{B}}}
\newcommand{\muts}    {\ensuremath{\mu_{[3.6\mm]}}}
\newcommand{\obs}     {\textrm{obs}}
\newcommand{\Rd}      {\ensuremath{R_\textrm{d}}}
\newcommand{\Risotf}  {\ensuremath{\textrm{R}_{25}}}
\newcommand{\sigln}   {\ensuremath{\sigma_\textrm{orth}}}

\newcommand{\Vd}      {\ensuremath{V_\textrm{d}}}
\newcommand{\Vf}      {\ensuremath{V_\textrm{f}}}
\newcommand{\Wmax}    {\ensuremath{W_\textrm{max}}}

\newcommand{\I}       {\ensuremath{I}}
\newcommand{\magas}   {\ensuremath{\textrm{ mag arcsec}^{-2}}}
\newcommand{\Csb}     {\ensuremath{\textrm{C}_\textrm{sb}}}

\begin{document}
\title[Scaling relations]{Scaling relations of mass, velocity and radius for disk galaxies}

\author{Earl Schulz}
\affil{60 Mountain Road, North Granby, CT 06060} \email{earlschulz@gmail.com}

\begin{abstract} 
  I demonstrate four tight correlations of total baryonic mass, velocity and radius for a set of nearby disk galaxies: the Mass-Velocity relation $ \Mt \propto V^4$; the Mass-Radius relation \hbox{$ \Mt \propto R^2$}; the Radius-Velocity relation $R \propto V^2$; and the Mass-Radius-Velocity relation $ \Mt \propto R V^2$. The Mass-Velocity relation is the familiar Baryonic Tully-Fisher relation(BTFR) and versions of the other three relations, using magnitude rather than baryonic mass, are also well known. These four observed correlations follow from a pair of more fundamental relations. First, the centripetal acceleration at the edge of the stellar disk is proportional to the acceleration predicted by Newtonian physics and secondly, this acceleration is a constant which is related to Milgrom's constant. The two primary relations can be manipulated algebraically to generate the four observed correlations and allow little room for dark matter inside the radius of the stellar disk. The primary relations \emph{do not} explain the velocity of the outer gaseous disks of spiral galaxies which \emph{do not} trace the Newtonian gravitational field of the observed matter.
\end{abstract}
\section{Introduction} \label{sect:Intro}
\subsection{Background}
Galaxy scaling law relations of mass, rotational velocity and radial size reveal the laws which govern the structure and dynamics of disk galaxies. Power-law scaling relations have been found which relate these parameters to the others in every combination. Due to observational constraints, scaling laws have been defined using a variety of proxies for the most fundamental properties. Magnitude or central brightness is used as a proxy for mass, various size measures(scale length, half-light radius, etc.) are used and the rotational velocity is measured in diverse ways.

The Tully-Fisher Relation (TFR) of disk galaxies relates luminosity to circular velocity \citep{Tul77} and, being a key component of the distance ladder, has been the subject of many studies of the effects of different wavelengths, galaxy morphology and environments\citep[e.g.,][]{lag13,moc12,nei14}. The fit of the TFR has been continuously improved by the availability of large surveys and improved distance measures. The scatter of the TFR is significantly smaller in the (NIR) band than in the UV or FIR bands and because near infrared luminosity is the best measure of stellar mass this suggests that the TFR is primarily a relation between mass and rotational velocity. Accordingly, the BTFR \citep{McGau00::1,Fre99,McGau05::1} correlates total baryonic mass including both the stellar and gas mass to \Vf, the `flat' rotational velocity measured in the outer \HI\ regions of disk galaxies resulting in a tight relationship over a range of many decades of mass.

Because of it's fundamental importance, the TFR has been the subject of many evaluations over a period of 40 years.  \cite{Bot87} and \cite{Mou89} provide usedful background and \cite{Bot87} was one the first to study the uncertainties of the I-band TFR which are significantly smaller than in other wavelengths.  \cite{Sak00} includes a definitive study of the random and systematic errors of the TFR calibration performed as part of the Hubble  Key Project.  \cite{Cou14} is a recent review of the reliability of the various methods used to measure galaxy mass.

The relations of mass, radius and velocity other than the TFR receive less attention but are equally important in helping us to uncover the underlying physical processes. \cite{Sai11} presents correlations among the parameters of luminosity, velocity and size equivalent to those given here but with much larger scatter. \cite{Sai11} makes several findings related to the source of errors in the measurements finding, in particular, that using an isophotal radius rather than a scale length reduced the scatter of the R-L relation by a factor of about 3 so that an isophotal radius is the more reliable measure.

The interpretation of scaling laws is tied to the dark matter paradigm which underpins the accepted \LCDM\  cosmology. \cite{gar11} is a current review of the subject and \cite{san10} is a history which gives invaluable background. The original puzzle was that the velocity of the stellar disk is often flat from the innermost regions to the edge of the optically observed disk and this was taken to mean prima facie that a dark matter halo is required to explain the kinematics. However \cite{kal83::2}, \cite{vanAlb85} and others showed that the combination of a central bulge surrounded by a truncated exponential disk results in a flat velocity curve with a reasonable stellar mass-to-light ratio and this finding is consistent with analytical results of Newtonian dynamics \citep{Mes63,Sch12}.

Maximum disk disk models \citep[see for example][among many others]{Beg87,Sel88,Pal00} find that little or no dark matter is required to explain the rotation curve in the inner regions of disk galaxies(i.e. the first few scale lengths) although others \citep[e.g.][]{cou99,ber11} find that dark matter is dominant at all radii. Some investigators force the result by adjusting the \ML\ ratio so that the contribution from baryons is small enough to `leave room' for an isothermal dark halo.

A purely exponential disk has $V_{max}$ at 2.2 scale lengths and most optical rotation curves  did not extent out much beyond that radius.  It is necessary to measure the rotation curve beyond  about 3 scale lengths (which contains 90\% of the disk light) to observe `flat' rotation curves. 
Observations of neutral hydrogen extending well beyond the optical radii of galaxies  proved beyond doubt that the rotation curves of the outer regions of spiral galaxies cannot be explained by the observed matter  \citep[e.g.:][]{Bos81::1,Bos81::2,Fre70,vanAlb86,Beg89,Beg91}.

Here I show that the dynamics of the inner stellar disk can be explained by Newtonian physics without room for a dominant amount of dark matter. This conclusion is robust because the uncertainties involved in the present calculation of average quantities are significantly smaller than those associated with previous evaluations.

I use the parameter \Rd\ which is the most meaningful measure of stellar disk size for the study of the dynamics.  \Rd\ is the radius at which the collapsed surface density, \ML, is small enough that it does not contribute significantly to the gravitational field.  The average mass density of a the stellar disk  is typically about 75  \Msun/\Lsun and I take \Rd\ to be the radius at which \ML\ decreases to below about 5\% of this value. 
\subsection{The Thesis}
Power-law scaling relations define planes in log space. Elementary geometry requires that if a set of galaxies lie on each of two planes then all of the galaxies must lie on the line at the intersection of the planes and, indeed, several observers have pointed out that the scaling relations of disk galaxies seem to have only one degree of freedom.

Any of the planes formed by rotation around the line of intersection also define power law relations which are as valid as any other. Any two relations of the three variables are sufficient to define the line of intersection and with this simple geometric insight I identified the physically meaningful relationships which define the line of intersection.

The equalities which govern the relationships of total baryonic mass, rotational velocity and disk radius are:
\begin{equation} \label{equ:govern}
   \frac{ \Cf\ \G\ \Mt}{\Rd^2} = \frac{\Vd^2}{\Rd} = \Aedg ,\end{equation}
where
\Mt\ is the total baryonic mass and is the sum of \Ms, the stellar mass, and \Mg, the gas mass;
\G\ is the the gravitational constant;
\Rd\ is the radius of the stellar disk;
\Vd\ is the circular velocity measured at \Rd;
and \Cf\ and \Aedg\ are constants which have physical interpretations.

Algebraic manipulation of Equations \ref{equ:govern} produces the Mass-Radius-Velocity relation, the Radius-Velocity relation, the Mass-Radius relation and the Mass-Velocity relation as given in Table \ref{tbl:fundrel}.

In what follows I use a well-characterized data base to find the constants \Cf\ and \Aedg\ and show that these two constants alone determine the four scaling relations with no need for any other free parameters.
\subsection{The Line of Intersection}
The scaling relations are defined over the log space ${\bigl<\log(\Mt), \log(\Rd), \log(\Vd)\bigr>}$. The line of intersection in this log space of the planes given in Table \ref{tbl:fundrel} is given perametrically as
\begin{equation} \label{equ:linedef}
    {\Lvec = \Lzvec +\textbf{w} t}\end{equation} where
$\Lzvec $ is any point on the line,
\begin{equation}
\textbf{w}=\frac{1}{\sqrt{21}}\bigl(4,2,1\bigr) \end{equation}
 is the unitary direction vector and t is the parameter.

Any plane of rotation around \Lvec\ can be specified by a equation with the form ${\textbf{v} \cdot \textbf{w} = 0}$ where ${\textbf{v}  = (a,b,c)}$ gives the exponents of a plane
\begin{equation} \label{equ:planedef}
M^a\times R^b\times V^c=constant
\end{equation}

which must satisfy $$4a+2b+c = 0.$$ Each of the relations given in Table \ref{tbl:fundrel} satisfy this equality as do an infinite number of other relations of mass, radius and velocity.

A value of \Lzvec\ can be found most straightforwardly by noting that the least squares regression line to a set of data passes through the centroid of the data where the centroid is the point consisting of the average of each of the components. In what follows \Lzvec\ is set to the centroid of the data and this fully specifies the line \Lvec\ and the constants \Cf\ and \Aedg.
\begin{deluxetable}{l}
\tablecolumns{1}
\tablewidth{0pt}
\tablecaption{Relations of M, V \& R}
\startdata
\sidehead{\textit{Fundamental Relations}}
$\displaystyle \frac{\Vd^2}{\Rd}       =\Cf \frac{\Mt \G}{\Rd^2} = \Aedg$ \phm{Phantom string for spacing. xxxx }\\[0.3cm]
\sidehead{\emph{Observed Correlations}}
$\displaystyle \Mt  = \frac{1}{\Cf \G}       ~\Rd \Vd^2   $ \\[0.3cm]
$\displaystyle \Rd   = \frac{1}{\Aedg}        ~\Vd^2       $ \\[0.3cm]
$\displaystyle \Mt   = \frac{\Aedg}{\Cf \G}   ~\Rd^2       $ \\[0.3cm]
$\displaystyle \Mt   = \frac{1}{\Cf \Aedg \G} ~\Vd^4       $
\enddata
\tablecomments{Here \Rd\ is the radius of the stellar disk, \Vd\ is the velocity at the edge of the stellar disk, and \Mt\ is the total baryonic mass. The constants \Cf\ and \Aedg\ are determined from the centroid of the data.}
\label{tbl:fundrel}
\end{deluxetable}
\begin{deluxetable*}{rrrccccrr}
\tablecolumns{9}
\tablewidth{0pc}
\tablecaption{Data for  \cite{McGau15} galaxy sample}
\tablehead{ \colhead{Galaxy }&\colhead{D }&\colhead{\Vd }&\colhead{Ref\tablenotemark{a}}&
 \colhead{\Ms  } & \colhead{\Mg }& \colhead{\Mt}&\colhead{R[3.6]\tablenotemark{b}}&\colhead{\Rd }\\
 & \colhead{Mpc} &\colhead{\kms}& \colhead{}& \colhead{\Msun}& \colhead{\Msun}& \colhead{\Msun}& \colhead{arcmin}&\colhead{kpc}   }
\startdata
      DDO154 &  4.04 &      53 &1,2 &  4.35E+07 & 1.38E+08 & 1.82E+08 &  1.080 &   1.27 \\
      D631-7 &  5.49 &      53 &1,2 &  5.84E+07 & 2.24E+08 & 2.82E+08 &  0.900 &   1.44 \\
      DDO168 &  4.25 &      53 &1,2 &  1.13E+08 & 4.48E+08 & 5.61E+08 &  1.700 &   2.10 \\
      D500-2 & 17.90 &      68 &1   &  2.37E+08 & 1.23E+09 & 1.47E+09 &  0.600 &   3.12 \\
     NGC2366 &  3.27 &      60 &1   &  4.31E+08 & 1.01E+09 & 1.44E+09 &  2.450 &   2.33 \\
      IC2574 &  3.91 &      77 &3   &  1.36E+09 & 2.11E+09 & 3.47E+09 &  3.200 &   3.64 \\
      F563-1 & 52.20 &     111 &1   &  4.31E+08 & 5.03E+09 & 5.46E+09 &  0.415 &   6.30 \\
     NGC2976 &  3.58 &      86 &1   &  1.91E+09 & 3.57E+08 & 2.27E+09 &  2.605 &   2.71 \\
     F568-V1 & 84.80 &     124 &1   &  2.52E+09 & 5.69E+09 & 8.21E+09 &  0.300 &   7.40 \\
     NGC1003 & 10.20 &     114 &1   &  3.33E+09 & 7.77E+09 & 1.11E+10 &  2.205 &   6.54 \\
      F568-1 & 95.50 &     116 &3   &  4.23E+09 & 5.83E+09 & 1.01E+10 &  0.350 &   9.72 \\
     NGC7793 &  3.61 &     110 &2   &  4.55E+09 & 1.46E+09 & 6.01E+09 &  4.310 &   4.53 \\
      UGC128 & 58.50 &     131 &1   &  6.70E+09 & 9.70E+09 & 1.64E+10 &  0.355 &   6.04 \\
     NGC2403 &  3.16 &     136 &1   &  7.28E+09 & 4.26E+09 & 1.15E+10 &  6.250 &   5.75 \\
      NGC925 &  9.43 &     114 &1   &  9.87E+09 & 6.74E+09 & 1.66E+10 &  3.550 &   9.74 \\
     NGC3198 & 13.80 &     149 &1,2 &  1.88E+10 & 2.01E+09 & 2.08E+10 &  3.510 &  14.09 \\
     NGC3621 &  6.56 &     152 &3   &  1.95E+10 & 1.97E+10 & 3.92E+10 &  4.250 &   8.11 \\
     NGC3521 &  8.00 &     192 &2   &  5.63E+10 & 8.58E+09 & 6.48E+10 &  4.305 &  10.02 \\
     NGC3031 &  3.65 &     199 &1   &  6.11E+10 & 1.20E+10 & 7.31E+10 & 10.650 &  11.31 \\
     NGC5055 &  8.99 &     181 &2   &  8.21E+10 & 1.34E+10 & 9.55E+10 &  6.840 &  17.89 \\
     NGC2998 & 68.30 &     212 &3   &  9.60E+10 & 4.90E+10 & 1.45E+11 &  1.420 &  28.21 \\
     NGC6674 & 51.90 &     241 &1   &  1.39E+11 & 6.65E+10 & 2.05E+11 &  1.750 &  26.42 \\
     NGC7331 & 14.90 &     246 &2   &  1.56E+11 & 1.92E+10 & 1.76E+11 &  5.830 &  25.27 \\
      NGC801 & 75.30 &     219 &2   &  1.61E+11 & 6.82E+10 & 2.29E+11 &  1.635 &  35.81 \\
     NGC5533 & 59.40 &     240 &2   &  1.83E+11 & 5.24E+10 & 2.35E+11 &  1.930 &  33.35 \\
     UGC2885 & 75.90 &     298 &1   &  2.82E+11 & 7.03E+10 & 3.52E+11 &  1.535 &  33.89 \\
\enddata
\tablenotetext{a}{References for Distance and velocity are  1)\cite{beg08}; ~2)\cite{sta09}; 3)\cite{tra09}. }
\tablenotetext{b}{Linearly projected angular size as described in the text.}
\label{tbl:TheData}
\end{deluxetable*}
\subsection{Organization}
This paper is organized as follows:
 Section 2 discusses the radii of disk galaxies;
 Section 3 describes the galaxy sample and motivate the definitions of the parameters \Mt, \Vd~ and \Rd;
 Section 4 presents the results of fitting the scaling laws;
 Section 5 compares the results to previous work;
 Section 6 summarizes.
\section{The Size of Disk Galaxies}\label{sect:DiskTrunc}
Correlations of mass, rotational velocity and radius reflect dynamic relationships and the definitions of the parameters should be consistent with the physics which tie them together. The parameters of total baryonic mass and ``final velocity" are unequivocal and  result in the Baryonic Tully-Fisher relation with very small dispersion.

Defining a physically meaningful radius of a disk galaxy has proven to be more difficult because the massive dark matter halo is thought to be entirely dominant in the outer regions and because the surface mass density of \HI\ gas  often decreases exponentially to distances far beyond the visual boundary of the stellar disk making it seem that disk galaxies do not have unique radii.

It was long thought that stellar disks extended great distances with a single exponential scale length\citep{Fre70} and this understanding lead to the frequently used ``maximum disk" methodology to fit rotation curves. \cite{Big10,Des14} found that the stellar disk extends outward far beyond \Risotf\ but at very faint integrated magnitude. Star formation efficiency is much lower in the outer disk and the Kennicut-Schmidt relation displays a change of exponent near \Risotf.  The most accurate observations use sensitive wide-area star counts to determine surface mass density and find large scale structure extending as far as 3 to 4 times the \Risotf\ radius.  It could be argued that the observed size of the stellar disk is determined by measurement sensitivity and harassment by companions rather than being a fundamental property of the galaxy.

\cite{vanderKru07} studied 23 edge-on spiral galaxies in the Sloan Digital Sky Survey and the STScI Digital Sky Survey and found that the majority of the galaxies show clear evidence for complete truncation, consistent with previous findings in other samples. In many cases, the truncation radius was associated with a drop in rotational velocity and with large \HI\ warps, starting at about 1.1 truncation radii. These truncations were predicted by \cite{Cas83} who showed that a truncated exponential disk could reproduce the rotation curves of most disk galaxies.

Various studies of inclined galaxies have not confirmed the complete truncations seen for edge-on galaxies. Instead, the stellar mass distribution follows a smooth exponential to a breakpoint where: 1) Some galaxies continued exponentially with the same scale factor; 2) Some(called truncated) continue but with a smaller scale factor; and 3) Some(called anti-truncated) continue but with a larger scale factor \citep[see e.g.,][]{erw05,poh06,poh08}.  \cite{Mar14} has now resolved the dichotomy between the observations of edge-on and inclined galaxies by showing that the stellar haloes of disk galaxies can outshine the brightness of inclined galaxies near the edge of the disk.  Not surprisingly, the density of the stellar halo falls off exponentially but the scale factor is, in general, different from the scale factor of the disk proper.

\section{The Data}\label{sect:TheData}
\subsection{Recent Measurement Improvements}
Recent work has reduced the dispersion of the scaling laws of mass, size and velocity dramatically so that the way that the scaling relation are related can now be seen. The improvements are the result of refinements to the measurement of each of the variables, including changes to the definition of what is measured.
\begin{itemize}
\item  Distance measurements using the Hubble law are affected by peculiar motions which results in errors of as much as 2 for nearby galaxies. Doppler independent  measurements now provide distance measures which are accurate to a range of about 5 to 10\%.
\item The mass-to-light ratio used to convert visual magnitude to mass was uncertain by about a factor of about 2. The corresponding uncertainty for recently available NIR measurements is much smaller.
\item The error due to  absorbtion by dust in the visual bands can be as large as a factor of 2 to 3 in the case of an edge-on spiral. The corresponding uncertainty for NIR measurements is negligible.
\item The Baryonic Tully-Fisher Relation \citep{McGau05::1} uses total baryonic (stellar mass plus gas mass) mass rather than only stellar mass and this modification to the definition of the  TFR reduces the the dispersion of the Tully-Fisher relation by \app2 for low-mass galaxies.
\item  The  error in measuring circular velocity from \HI\ bandwidths can be as large as 40\% and this error translates to an error as large as a factor of a few in the Tully Fisher relation due to the  V$^4$ dependence. \cite{McGau15}  used  the `flat' asymptotic outer gas velocity fit to tilted ring model rather than the velocity derived from a line width and the dispersion of the resulting velocity measurement is less than 10\%.
\item A contribution of the current work is to use \Rd, the linearly projected edge of the stellar disk seen in the 3.6 \mm\ band, as the measure of galaxy size and this results in dramatically smaller dispersion of the relations involving galaxy size.
 \end{itemize}
\subsection{The Galaxy Sample}
I use the set of well-observed nearby galaxies chosen by \cite{McGau15} which consists of galaxies with photometry in the optical and in the \emph{Spitzer Space Telescope} 3.6 \mm\ band and which have an high quality extended rotation curves from 21 cm interferometer measurements. The selection included only galaxies which have `flat' rotation curves in the outer HI disk and an accurate velocity curve which increases to a constant velocity. Galaxies which showed a significantly decreasing velocity at any radius were rejected.
Also, the selection was limited to galaxies with `consistent' velocity curves
by rejecting galaxies for which the \HI\ curve showed excessive turbulence or for which the \HI\ disk was misaligned relative to the stellar disk.
The  sample consists of 26 galaxies. Of these, the smaller galaxies are almost entirely Sd and Sm spirals and irregulars with rotation velocities  $\Vd < 90~\kms$ and the larger galaxies are earlier types with \Vd\ as large as 450 \kms. The smaller galaxies are distinctly gas dominated such that the fraction of gas \Mg/\Mt\ is as large as 80\% whereas the fraction is in the range of approximately $10\% < \Mg/\Mt < 30\%$ for the larger galaxies.
\subsection{Circular Velocity}
The Tully-Fisher relation originally used \HI\ bandwidth, defined to be full width at 20\% maximum flux corrected for inclination, as a measure of circular velocity. \cite{Tul77}  noted that the slope and scatter of the relation are sensitive to the exact definition of the velocity term. \cite{Cou09} and \cite{Cou11} investigated various measures and found that their parameter \Wmax\ reduced the rms dispersion by about half. \Wmax\ is based on the \HI\ bandwidth at 50\% maximum within the band enclosing 90\% of the total integrated flux, excluding the problem of `wings' caused by high velocity in the central regions of some bright galaxies.

Here I adopt the \cite{McGau15} terminal velocity \Vf\ values without change where \Vf\ is the  asymptotic, nearly constant, velocity of the outer \HI\ gas  corrected for inclination and rejecting galaxies which do not rise smoothly to a constant velocity. The scatter involved in measuring \Vf\ is smaller than for \Wmax\ although at the cost of limiting the sample selection.

For these galaxies, the change from an increasing to a flat velocity occurs occurs at approximately the optical radius so that the maximum velocity, the `flat' velocity \Vf, and the velocity at edge of the stellar disk \Vd\ are equal to within the uncertainties of the data.

\begin{figure}[t]
   \plotone{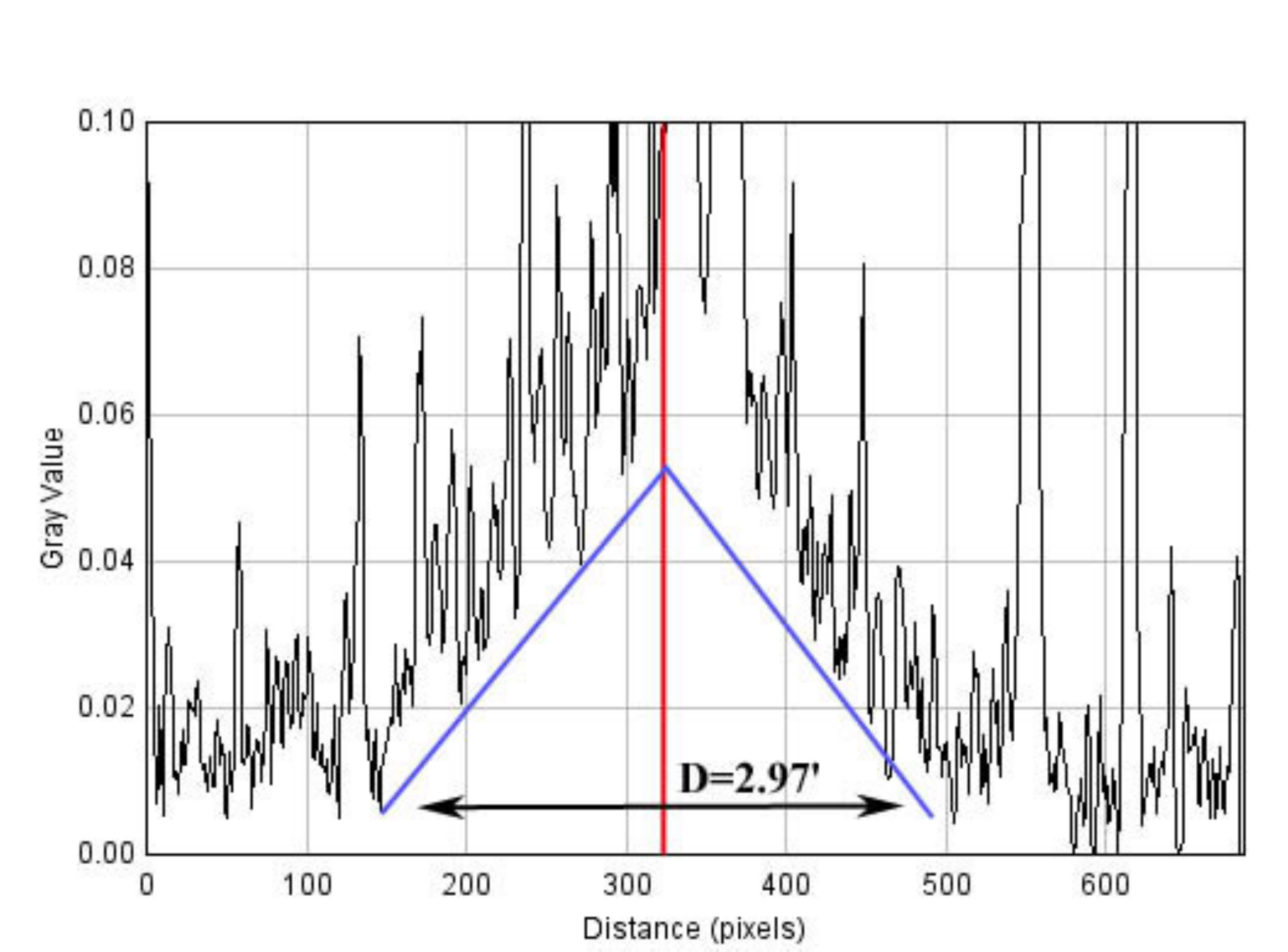}
   \caption{Flux trace along the major axis of DDO 168. Blue lines are symmetric bounds projected linearly to find the limit at which the flux decreases to the local background level.  The linearly projected edge can be determined to within about $\pm10\%$  despite an extremely noisy field. }
   \label{fig:DDO168}
\end{figure}
\begin{figure}[t]
   \plotone{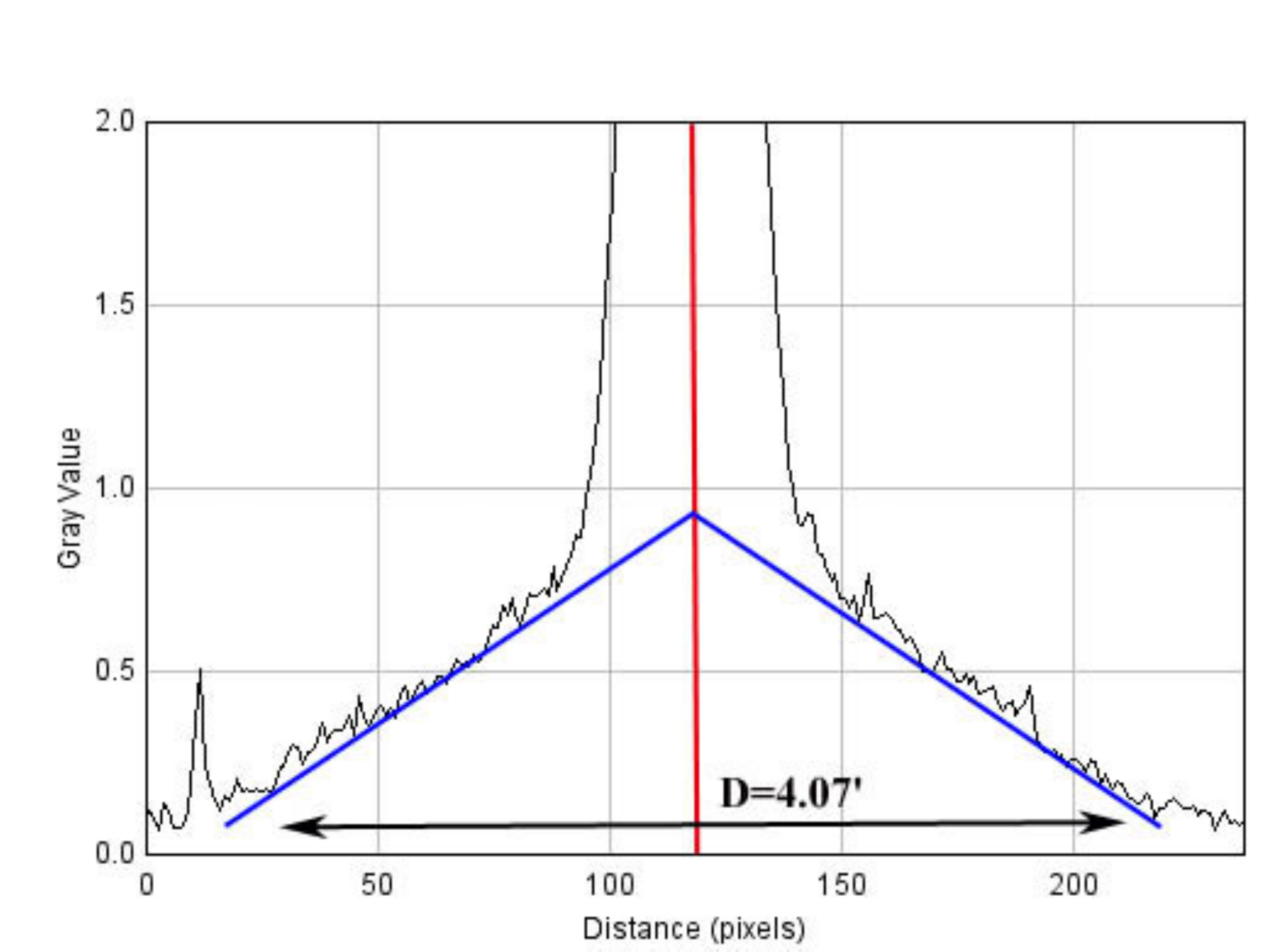}
   \caption{Flux trace along the major axis of NGC 936.  The linearly projected edge is a more meaningful measure of the size of the stellar  compared to than $R_{\onehalf}$ or an exponential scale factor which measures the size of the innermost galaxy. Although NGC 936 is not part of the current data set it is presented for comparison to
   Figures 1 and 6 of \cite{Mun15}.}
   \label{fig:N936}
\end{figure}
\begin{figure}[t]
   \plotone{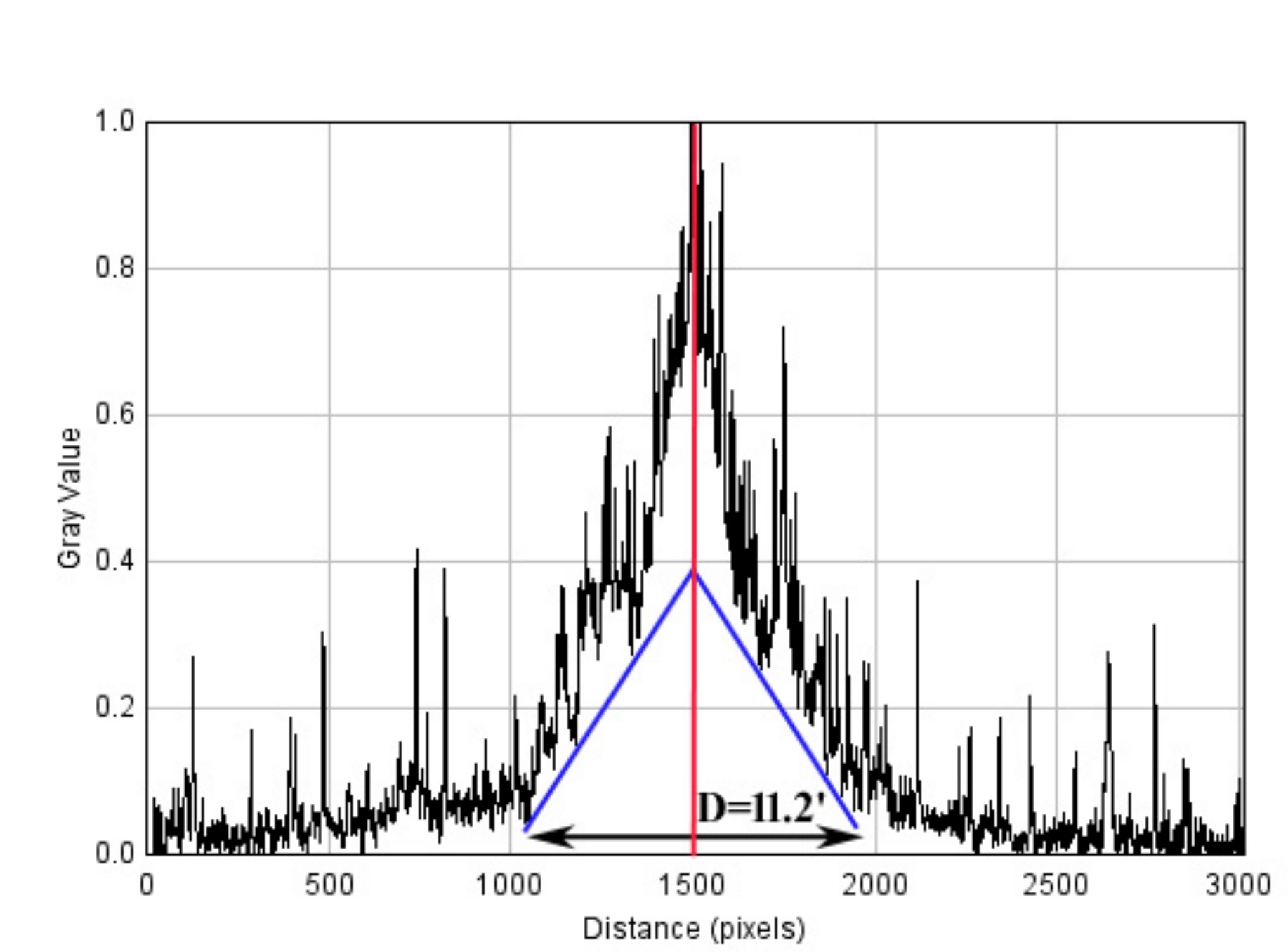}
   \caption{Flux trace along the major axis of NGC 300.  The linearly projected edge measures the boundary at which the gravitational contribution of baryonic mass becomes negligible and this is the size measurement which minimizes the scatter of the scaling relations.  The disk of NGC 300 is observed to extend far beyond this boundary.}
   \label{fig:N300}
\end{figure}

\subsection{Total Baryonic Mass}
The total mass of the stellar disk, \Mt, is  the sum of the gas mass \Mg\ and the stellar mass \Ms.
\HI\ mass is taken from \cite{McGau15} where it is determined  from the integrated 21 cm line flux as $\MHI=2.36\dex{5} D^2F_{HI}$ and total gas mass is $\Mg = \eta \MHI$ where the factor $\eta=1.4$ accounts for helium and \Htwo\citep{gur10}.
Stellar mass was found by multiplying the value of [3.6] luminosity given in \cite{McGau15} by the stellar mass-to-light ratio \MLnir=0.6 recommended by \cite{Mei14}.  Note that this differs from the  value  of \MLnir=0.47 used by  \cite{McGau15} which was determined by requiring ``photometric self-consistency" of the population synthesis models.  \cite{Mei14} found that the error in \Ms\ resulting from a constant value of \MLnir=0.6 is less than about 0.1 dex (\app 15\%). \cite{Nor14} independently confirmed this result  which is a great improvement compared to B-band observations for which the error caused by an uncertain \ML\ can be on the order of a few. The   \cite{Mei14} results are compelling but  contrary results should be kept in mind. Table 4 of \cite{McGau15} compiles various values of \MLnir ranging from 0.14 to 0.6 and  also  \cite{Que15} investigates the effect of dust contamination  which increases the uncertainties associated with \MLnir\ for dusty galaxies.

\subsection{The size of a Stellar Disk}\label{sect:RadiusDiscuss}
The size of stellar disks can be approximated by the radius of the \muB = 25 isophote corresponding to a mass surface density of \app9 \Msun pc$^{-2}$ or by the ${\muts = 27}$ isophote which corresponds\footnote{        
   Here brightness in \magas\  is converted to surface brightness \I\ in units of L0 pc$-2$ with the  equation:
   $$ 2.5\log(\I)  = -\mu + \Msun +\Csb $$
   where \Msun\ is the absolute solar magnitude in the same band as $\mu$  and ${\Csb=5\log( 180\times60\times60/(10\pi))=21.572}$ is independent of wavelength and distance. Surface mass density is then found by multiplying by $\Upsilon$.  For the B-band:
   \Msun= 5.48(B Mag) and $\Upsilon_\star$ = 1.39\citep{Fly06}.
   For the [3.6] band; \Msun= 6.019(AB Mag) based on \Msun = 3.24(Vega) per \cite{Oh08} and $\Upsilon_\star$=0.6 as above.}
 to a surface density of \app1 \Msun pc$^{-2}$.
 However defining the radius to be some faint isophotal is problematic because the measurement then depends critically on an accurate determination of the background and because it's difficult to account for the effect of a stellar halo.  The alternatives of defining the size of a disk galaxy in terms of the half-light radius or a scale length also cause difficulties because these quantities are difficult to measure and are only proxies for disk size.

After some trial and error, I fixed on a simple way to measure the radius at which the linearly projected surface brightness drops below the local background. The projection uses the outermost region of the disk where the drop in surface brightness is approximately linear.  The intent of defining disk size in this way is to find the boundary of the gravitationally important baryonic matter.  Beyond the linearly projected radius the surface mass density so small that the disk no longer contributes to the gravitational attraction.
This definition is more robust than the radius of a faint isophotal because it does not depend on determining the exact global background level and is not sensitive to contamination by stars or background galaxies. The dispersion in finding the linearly projected edge is about 10--15\% which is small compared to using for instance \Risotf, the 25 \magas\ B-band isophote, for which independently determined values can vary by 50\% or more.

Figures \ref{fig:DDO168}-\ref{fig:N300} provide three examples which illustrate the advantages of using, \Rd, the linearly projected edge as a measure of disk size.

Figure \ref{fig:DDO168} shows the flux trace along the major axis of dwarf irregular DDO 168.  In this case \HI\ gas outweighs stellar mass by a factor of \app4 and stellar surface mass density is lower than for larger spirals by approximately the same factor. Even so, the linearly projected edge is well defined and finds the limit beyond which Newtonian physics fails to explain the rotation curve.

Figure \ref{fig:N936} shows the flux trace along the major axis of the barred lenticular NGC 936. The radial extent of NGC 936 in the 3.6 and 4.5 \mm\ band is analyzed in detail in \cite{Mun15} where it is shown that the 3.6 surface brightness declines smoothly to a radius of almost 200\arcsec\ where the brightness is only $\mu_{3.6} \sim 28$(AB) \magas. As shown in Figure \ref{fig:N936}, the linearly projected edge is \Rd=122\arcsec.  Beyond this radius the collapsed   surface density is too small to contribute appreciably to the rotation curve but is large enough to account for some star formation.  Figure 6 of \cite{Mun15} plots surface brightness versus radius out to the limit of detection and this plot makes it obvious why scale length is a poor measure of disk size.  Because  of the influence of the central bulge, scale length is small in the central region, much larger in the region from the bulge to \Rd, and is again smaller in the outermost region. The variation in scale length between regions is a factor of approximately 2-4 for this typical case.

Figure \ref{fig:N300} shows the flux trace along the major axis of NGC 300, a nearby very well observed Sc galaxy.  \cite{Hil16} recently presented the results of star counts  which show that the young stellar disk of NGC 300 is an unbroken  disk extending for 8 or even 10 scale lengths, consistent with \cite{Bla05}. 
NGC 300 may present a counter-example to the idea that disk galaxies are universally truncated but it does not invalidate the procedure to find \Rd\ given here.
\subsection{The Data}
Table \ref{tbl:TheData} presents the data used used in the present analysis:
 Column 1 gives the most common name of each galaxy;
 Column 2 is D, the distance to the galaxy(Mpc);
 Column 3 is the circular velocity, adjusted for inclination(\kms);
 Column 4 gives the original sources for the velocity data;
 Column 5 is \Ms, the stellar mass(\Msun);
 Column 6 is \Mg, the gas mass(\Msun);
 Column 7 is \Mt, the total disk mass(\Msun);
 Column 8 is R[3.6], the angular size of the stellar disk  as described in Section \ref{sect:RadiusDiscuss} and
 Column 9 is \Rd, the radius of the stellar disk(kpc).

\section{Results and Discussion} \label{sect:Results}
The centroid of the data is the vector of the  averages of the parameters and this vector is used to solve the system of equations given in Table \ref{tbl:fundrel} and find the constant \Aedg:
\begin{equation}
\log(\Aedg) = 2*\avg{\log(\Vd)} -\avg{\log(\Rd)}
\end{equation}
which evaluates to :
\begin{align}
\Aedg &= 2,052~\textrm{km}^2~\textrm{kpc}^{-1}~\textrm{sec}^{-2}         \nonumber \\
      &= 6.65 \dex{-11}\textrm{~m~sec}^{-2}\\
      &= 0.665 \textrm{~\AA~sec}^{-2}
\end{align}
and the constant \Cf:
\begin{equation}
\log(\Cf) = 2*\avg{\log(\Vd)} +\avg{\log(\Rd)} -\avg{\log(\Mt \G )}
\end{equation}
which evaluates to :
\begin{align}
\Cf &=  2.03.
\end{align}

The two constants \Aedg\ and \Cf\ are sufficient to generate the four scaling relations given in Table \ref{tbl:fundrel} as demonstrated by Figures \ref{fig:M-V} through \ref{fig:M-RV} below.
\subsection{The Mass-Velocity Relation}
\begin{figure}[t]
  \plotone{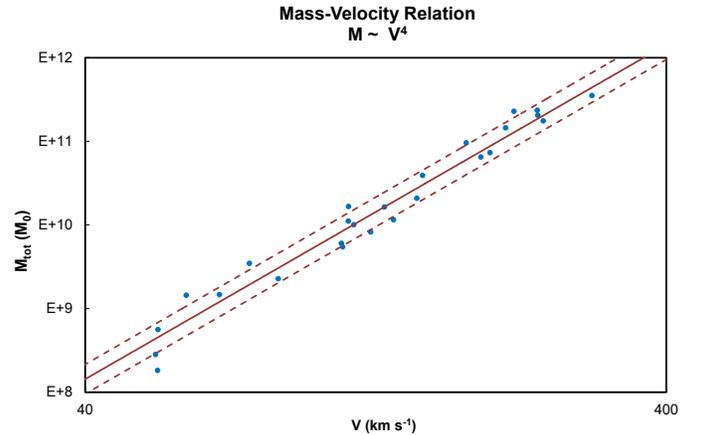}
  \caption{The Mass-Velocity Relation: $\Mt =  55.8 \Vd^{4}$.  }
    \label{fig:M-V}
\end{figure}The Mass-Velocity relation is
\begin{equation}\label{equ:M-V}
 \Mt = \frac{1}{\Cf \Aedg G} \Vd^{4}.
\end{equation}

which is
\begin{equation*}
 \Mt =  55.8 \Vd^{4}
\end{equation*}
in units of \Msun\ and \kms.

This is the Baryonic Tully-Fisher Relation and figure \ref{fig:M-V} is identical to that given in \cite{McGau15} except for a small offset due to using an updated \MLnir.

\subsection{The Mass-Radius Relation}
\begin{figure}[t]
  \plotone{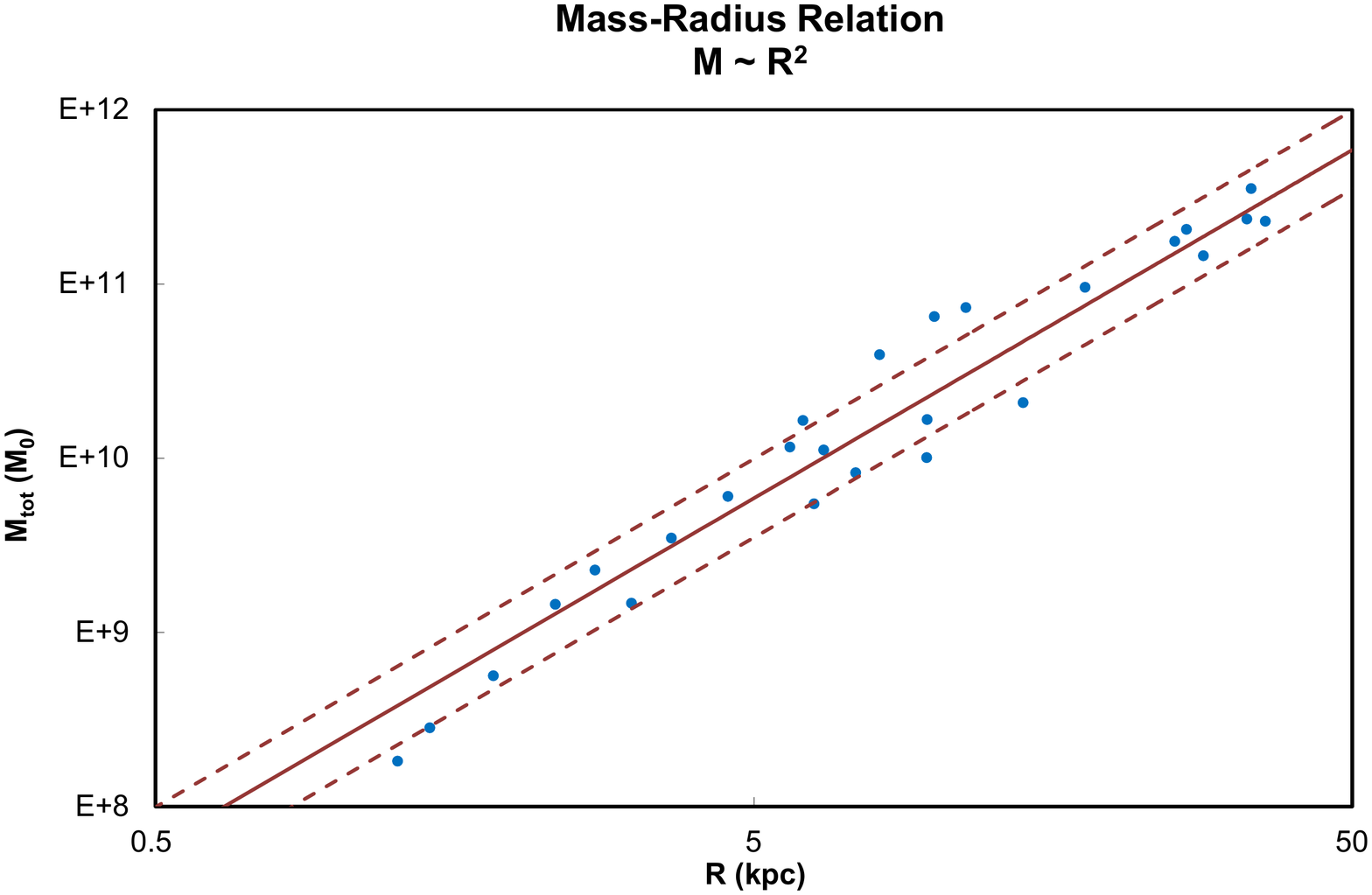}
  \caption{The Mass-Radius Relation: $\Mt = 2.35\dex{8} R^{2}$.}
  \label{fig:M-R}
\end{figure}
The Mass-Radius relation is shown in Figure~\ref{fig:M-R}:
\begin{equation}\label{equ:M-R}
 \Mt = \frac{\Aedg}{\Cf G} R^{2}
\end{equation}
which is
\begin{equation*}
 \Mt = 2.35\dex{8} R^{2}
\end{equation*}
in units of \Msun\ and \kpc.

The Mass-Radius relation is the same as the Luminosity-Diameter relation  of disk galaxies \citep{hei69,giu88,gav96}, a tight scaling relationship between the  magnitude and the diameter of the stellar disk, usually taken to be the A$_{25}$ isophote. More recently \cite{vdBergh08::1} found that the relation is independent of environment or local mass density, consistent with the conclusions of \cite{gir91}. \cite{vdBergh08::1,Mun15,Sai11},   did show a definite morphological dependance in that early-type galaxies are more compact than  late-type galaxies at the same luminosity.
\subsection{The Radius-Velocity Relation}
\begin{figure}[t]
  \plotone{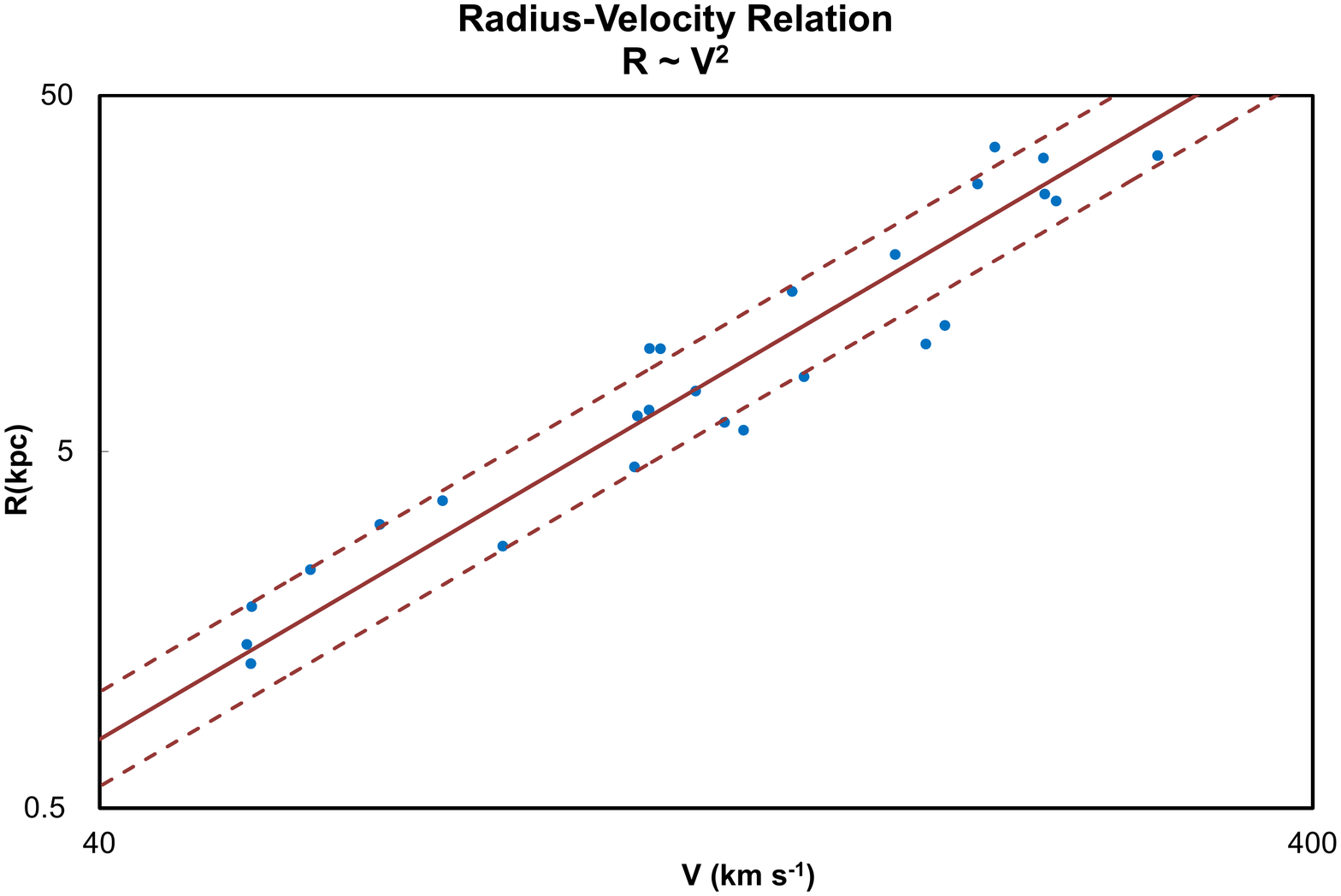}
  \caption{The Radius-Velocity Relation: $R = 4.87\dex{-4}~\Vd^{2}$ }
  \label{fig:R-V}
\end{figure}
The Radius-Velocity relation  shown in  Figure \ref{fig:R-V}:
\begin{equation}\label{equ:R-V}
 R = \frac1\Aedg \Vd^{2}
\end{equation}
which is
\begin{equation*}
 R = 4.87\dex{-4}~\Vd^{2}
\end{equation*}
in units of \kpc\ and \kms.
\subsection{The Mass-Radius-Velocity Relation}
\begin{figure}[t]
  \plotone{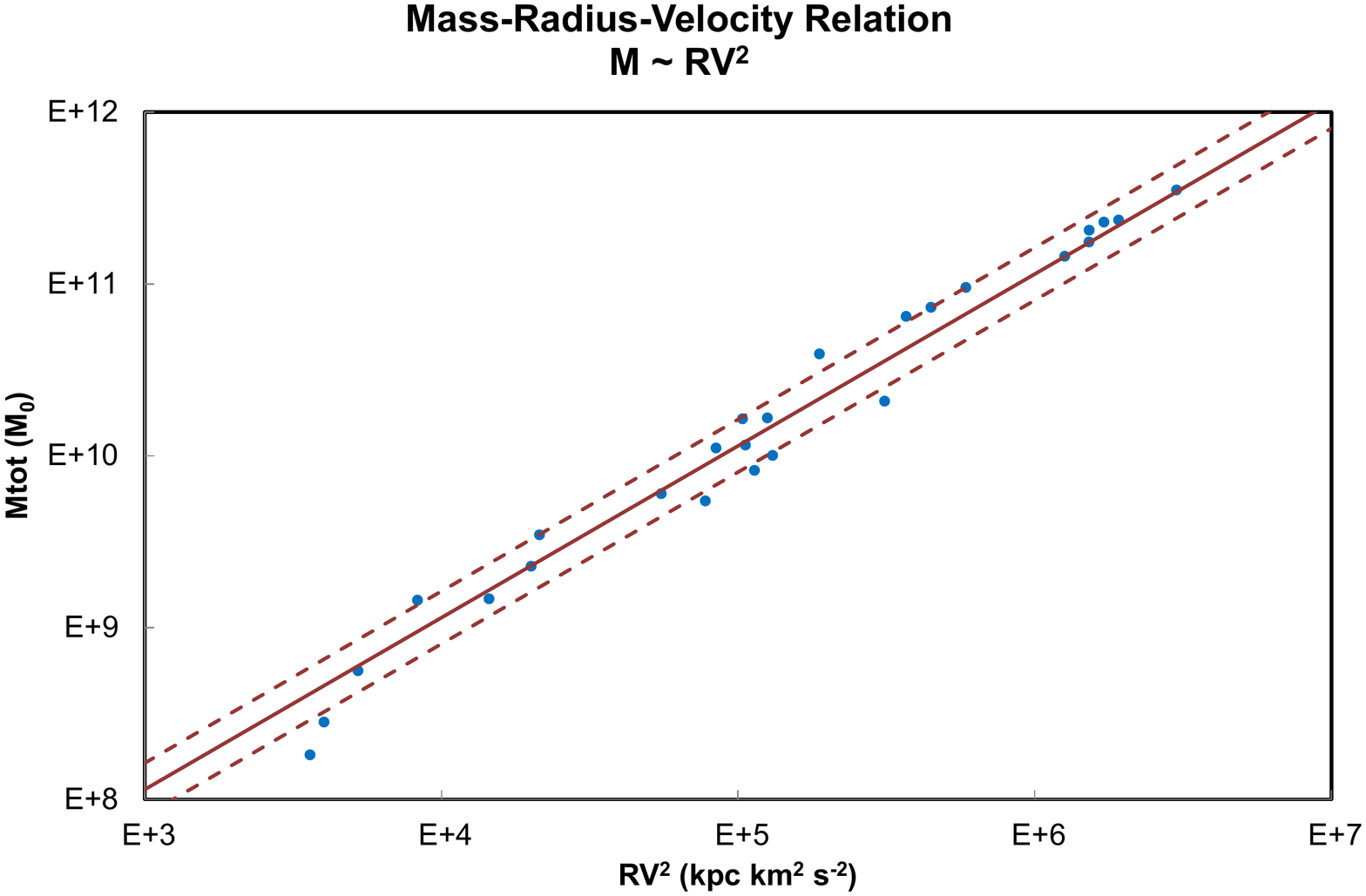}
  \caption{The Mass-Radius-Velocity Relation: $\Mt = 1.15\dex{5}~R \Vd^2 $}
  \label{fig:M-RV}
\end{figure}
The Mass-Radius-Velocity  relation is shown in Figure~\ref{fig:M-RV}:
\begin{equation}\label{equ:M-RV}
 \Mt = \frac{1}{\Cf G} R \Vd^2
\end{equation}
which is
\begin{equation*}
 \Mt = 1.15\dex{5}~R \Vd^2 
\end{equation*}
in units of \Msun, \kpc\ and \kms.

The Mass-Radius-Velocity relation implies that the centripetal acceleration at the edge of a galaxy disk is proportional to the acceleration due to gravity. This is the ``fundamental" relation $L\propto \Vd^2 R$ given in \cite{han01} who searched several galaxy samples consisting of more than 500 spiral galaxies in several optical bands to find the relation which has the smallest dispersion of any relation of L, V and R
 \citep[see also][]{Tul77,Kod00::1,Cou07,Piz05,Sai11,Wil10}. Using the isophotal radii in NIR or optical wavelengths resulted in markedly smaller dispersion than using a disk scale length. The \cite{han01} galaxy sample included only late-type spirals and covered a smaller range of galaxy size than the present work. No morphological effects were noted in their relatively homogeneous sample.
\subsection{Uncertainties}
\cite{Akr96} discuss the complications involved in the linear regression analysis of astronomical data. In principle ordinary least squares(OLS) fits do not allow for measurement uncertainties on both the X and Y variables and in addition, OLS cannot address the case where  \emph{correlated} uncertainties affect both measurements so that the slope of the line determined by OLS can vary significantly depending on the choice of (X,Y). Finally, OLS does not allow for ``intrinsic scatter" in the data which is the case when some of the data points are true outliers or follow some undiscovered rule. These problems are most troublesome when measurement errors are large and can be avoided by reducing all measurement errors to a negligible size.

Here I use the orthogonal standard deviation in log space, \sigln, as the error measure which is a common compromise. \sigln\ is the rms value of the distance of the data to each of the four planes defined in Table \ref{tbl:fundrel} and is found using the elementary formula for the distance of a point $(p_1,p_2,p_3)$ to a plane  aX+bY+cZ+d=0:
$$Distance = \frac{ ap_1 + bp_2 + cp_2 + d}{\sqrt{a^2+b^2+c^2}}$$ and converting to the natural log so that \sigln\ is close to the relative error ${\delta Y}/{Y}$. In a 2-D plane the formula reduces to the formula for the orthogonal distance from a point to a line.  Defined this way, the value of \sigln\ is independent of the assignment of variables as (X,Y).

Table \ref{tbl:MyStats} gives fitting statistics for each of the four relations.  The Pearson r coefficient is a measure of how well the data can be fit by a straight line where a value of 1.0 signifies a perfect fit and a value of 0.0 implies no correlation.  The values found here indicate a very good linear fit is available with little need to look for secondary corrections.\footnote{Note again that this is a \emph{selected} database which predominantly includes gas-rich late-type galaxies so that morphological effects aren't expected here but do seem to be present in broader samples such as \cite{Cou07} discussed below.}  Table \ref{tbl:MyStats} gives the scatter \sigln\ using the integer slopes given in Table \ref{tbl:fundrel} and also for optimized slopes which minimize \sigln.  The optimized slope of the M-RV relation is the optimal coefficient $\gamma$ of the relation $M\varpropto\bigl(V^2 R\bigr)^\gamma$.  The optimized slopes do reduce the scatter slightly but at the cost of introducing a new free variable.

The expected dispersion of each of the four relations due to random independent variation of the measured variables is:
\begin{equation}  u^2_i =  c^2_{\textrm{1,i}} u^2(\Mt) +   c^2_{\textrm{2,i}}u^2(\Vd)   + c^2_{\textrm{3,i}}u^2(\Rd)         \label{equ:dispersion} \end{equation}
where $u^2_i$ is the square of the dispersion;
the sensitivity coefficients $c_{\textrm{n,j}}$ are the coefficients of \Mt, Vd, and \Rd\ for each of the four scaling relations;
and $u(\Mt),u(\Vd)$ and $u(\Rd)$ are the dispersions of the variables. It's possible to use the observed dispersions relative to the four observed scaling relations to back out estimates of the scatter of the individual variables.  A set of $\bigl[ u(\Mt), u(\Vd), \Rd \bigr]$  which is roughly consistent with the measured dispersions are:

~~~~~~~~$\begin{array}{l  }
   \textrm{u(\Mt)}   =  0.04  \\
   \textrm{u(\Vd)}   =  0.02  \\
   \textrm{u(\Rd)}   =  0.12  \\
\end{array} $

These quantitative measures of the scatter are useful in understanding the scaling relations and to compare data sets. These dispersions do not include not systematic errors which are difficult to quantify but might be significant or even dominant. For example, an error in the distance ladder or in \MLnir\ would affect all of the measurements of \Mt\ and \Vd\ without increasing the observed scatter.
\begin{deluxetable}{lllll}[t]
\tablecolumns{5}
\tablewidth{0pc}
\tablecaption{Fitting Statistics for Table \ref{tbl:TheData} Data}
\tablehead{ \colhead{Description}&\colhead{M-V}&\colhead{M-R}&\colhead{R-V}&\colhead{M-RV}} \\
\startdata
Pearson r Coefficient     & 0.98     & 0.97     & 0.96    & 0.99   \\
\multicolumn{4}{c}{ ~ }     \\
Integer Slope             & 4        & 2        & 2       & 1      \\
Integer Slope \sigln      & 0.099    & 0.233    & 0.137   & 0.144  \\
 \multicolumn{4}{c}{ ~ }     \\
Optimized Slope           & 4.22     & 2.20     & 1.98    & 1.06   \\
Optimized Slope \sigln    & 0.095    & 0.216    & 0.137   & 0.137
\enddata
   \tablenotetext{ }{The square of the r coefficient is a measure of how much of the variation in the data can be explained by a linear fit. }
\label{tbl:MyStats}
\end{deluxetable}
\subsection{The \Cf\ Constant}
Newtonian potential theory predicts that the gravitational force of a flattened disk is greater at the edge of the disk than the force due to a point mass. The size of the increase depends on the the details of the mass distribution and Table \ref{tbl:FlatFact} gives analytical values for a few simple models.

The \emph{Exponential Disk} \cite{Bin08} is often used to model the kinetics of disk galaxies. The model assumes that the disk is infinite in extent with a exponential surface mass distribution. The increase of force for this disk at the radius of maximum velocity is much smaller than for a finite disk; \Cf =0.83 because the mass of the disk beyond the point of maximum counterbalances the effect of the mass of the inner disk.

The \emph{Finite Mestel disk}\citep{Mes63,Sch12} includes a central singularity where the rotation curve steps to a value which is then constant to the edge of the disk. This behavior is typical of many Sc and Sd galaxies, including the Milky Way. The  analytical value of the increase of force for this disk relative to a point mass is \Cf=1.6.

The \emph{Maclaurin disk}\citep{Mes63,Sch09} is the most appropriate model for the Table \ref{tbl:TheData} sample because the velocity curve rises linearly to the edge of the disk for the galaxies in the sample. The analytical value of the force increase for this disk is \Cf=2.4 which is not too far from the  measured value of the flatness factor, \Cf = 2.03 found in section \ref{sect:Results}. Note that the Maclaurin disk model is basically the same as the truncated exponential disk model described by \cite{Cas83}.

In summary, the value of \Cf\ derived from observations is close to what is expected from theory.
\begin{deluxetable}{lll}[t]
\tablecolumns{3}
\tablewidth{0pc}
\tablecaption{The Flatness Factor \Cf\ for Several Disks}
\tablehead{ \colhead{Name}&\colhead{Surface Mass Density}&\colhead{Flatness Factor}} \\[+5.0ex]
\startdata
   Maclaurin disk     & $ \Sigma = \frac{3 \Mt}{2 \pi \Rd^2} \sqrt{1 - {r^2}/{\Rd^2}}           $&$ \Cf=\frac{3\pi}{4}=2.356   $ \\[+5.0ex]
   Finite Mestel Disk & $ \Sigma = \frac{\Mt}{2 \pi \Rd^2 } \, \arccos\left(\frac{r}\Rd\right)  $&$ \Cf =\frac{\pi}{2} = 1.571 $ \\[+5.0ex]
   Exponential Disk   & $ \Sigma = \frac{\Mt}{2 \pi h^2} \exp{-\frac{r}{h}}                     $&$ \Cf =0.83                  $ \\
\enddata
\tablenotetext{ }{Theoretical flatness factors(\Cf) for three models of galaxy disks which describe for the increased gravitational force due to a flattened mass distribution versus a point mass vary by a factor 2 to 3 in the implied mass. See \cite{Mes63,Bin08,Sch12}. }
\label{tbl:FlatFact}
\end{deluxetable}
\subsection{The \Aedg\ Constant and MOND}
The idea that the acceleration at the edge of the stellar disk is roughly constant is not new. \cite{Tul77} introduced the scaling relation between the full width of the \HI\ velocity profile  measured at 20\% of the maximum intensity versus the Holmberg diameter where the exponent of the relation  $R \propto V^\gamma$ is close to 2.0. The form of this relation implies that $V^2/R$, the centripetal acceleration at the edge of disk galaxies, is approximately constant. More recently, \cite{don09} and \cite{gen09} have come to the same conclusion. This characteristic of galaxies has not yet been explained completely although \cite{gen09} suggests possible explanations within the context of the \LCDM\  paradigm.

The constant \Aedg\ is clearly associated with the \aZ\ constant predicted by the Modified Newtonian Dynamics (MOND) hypothesis \citep{Mil83,Mil83::1,San02,McGau12}. MOND proposes an ``effective" force law to explain the observed flat rotation curves of disk galaxies. Accelerations greater than \aZ\ are governed by Newtonian physics so that and $V^2/R \propto \aN$ but accelerations $a\lt \aZ$ are governed by an ``effective" force law
\begin{equation}
{\aeff = \sqrt{\aN \aZ}}  ~~~ \textrm{for } \aN < \aZ
\end{equation}
where \aZ\ is thought to be a universal constant with a value of \aZ $\approx$1.2 \AA~s$^{-2}$ and \aN\ is the acceleration calculated using classical theory.
\begin{figure*}
  \begin{subfigure} [{The M-V relation overlayed on \cite{Cou07} data. }]
    {\includegraphics[width=.48\textwidth]{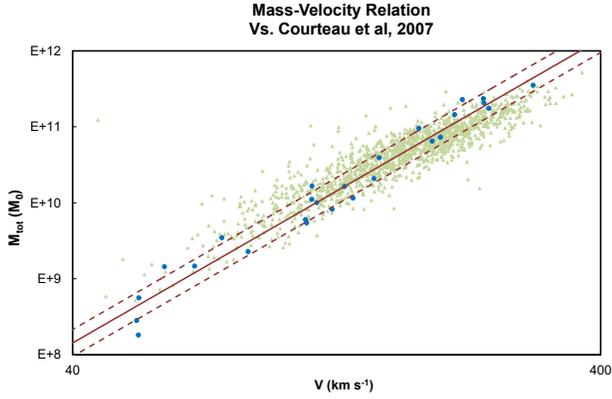}}
    \end{subfigure}%
  \begin{subfigure} [{The M-R relation shown in overlayed on \cite{Cou07} data. }]
     {\includegraphics[width=.48\textwidth]{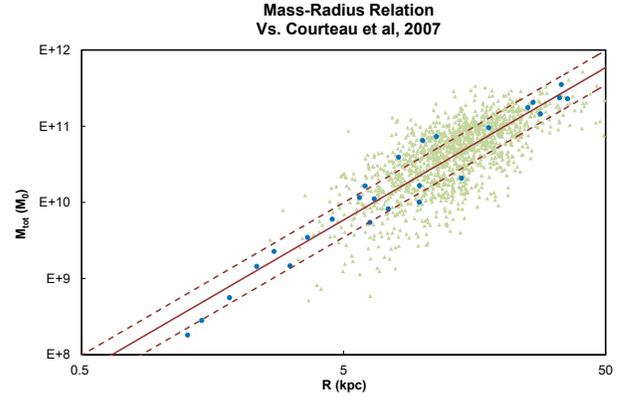}}
     \end{subfigure}
   \begin{subfigure}[{The R-V relation shown in overlayed on \cite{Cou07} data.  }]
     {\includegraphics[width=.48\textwidth]{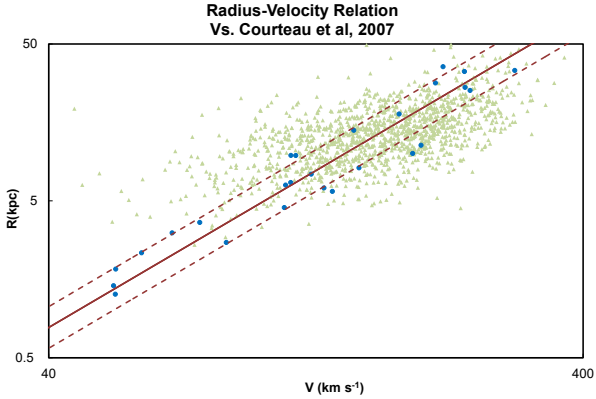}}
     \end{subfigure}%
   \begin{subfigure}[{The M-RV relation shown in overlayed on \cite{Cou07} data. }]
     {\includegraphics[width=.48\textwidth]{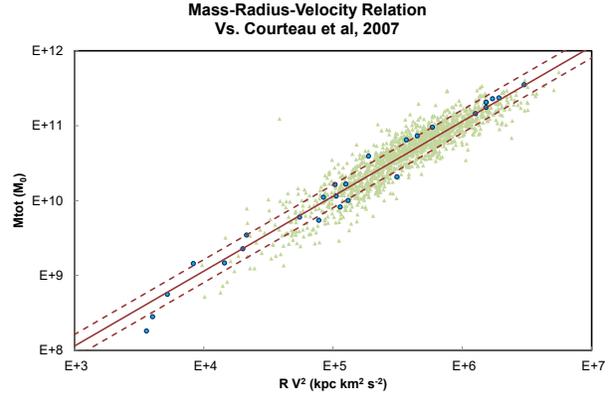}}
     \end{subfigure}
\caption{Comparison of the M, V, \& R Relations to \cite{Cou07}. The solid red line are the integer slopes given in Table \ref{tbl:fundrel} and the dashed red lines  are the $\pm 1 \sigma$ boundaries of the Table \ref{tbl:TheData} data.  }
\label{fig:Cou07}
\end{figure*}
At large distances from the galaxy, \aN\ approaches the Keplarian solution ${\aN = M G /R^2}$ and \aZ\ can be found from measured baryonic mass and terminal velocity:

\begin{equation}\label{equ:M-V-fromMOND}
\sqrt{\aZ}   = \frac{\Vf^2}{\sqrt{\Mt G}}~~~ \textrm{for \aN} \rightarrow 0
\end{equation}

In practice, \aZ\ is found by fitting a selection of galaxies to the BTFR (i.e.,the M-V) relation:
\begin{equation}\label{equ:M-V-fromMONDL}
\Mt = \frac{\lambda}{\aZ G } \Vf^4
\end{equation}
and as discussed in \cite{McGau12}, the parameter $\lambda$ is introduced to  account for the finite size and non-spherical geometry of galaxies measured at a finite radius.  Equation \ref{equ:M-V-fromMOND} has the same form as equation \ref{equ:M-V} where $\Cf$ is equivalent $1/\lambda$ and \Aedg\ to \aZ.

The values of the constants $\Cf$, $1/\lambda$, \Aedg\ and \aZ\ differ because \cite{McGau12} adopted a value of $1/\lambda=1.125$ which is smaller than the expected value of $1.5 < 1/\lambda < 2.5$ as given in Table \ref{tbl:FlatFact}.

One of the arguments in favor of  MOND has been that it eliminates the troubling `disk-halo conspiracy' implicit in assuming a CDM halo \citep{Bah85,vanAlb86}.  However, a transition is also required for MOND models and three forms of the empirical $\mu$ joining function are currently used \citep{Bug15}.  A large value of $1/\lambda$ would be a problem for MOND theory because it would be difficult to find a reasonable way to join the results for $a>>\aZ$ and $a<<\aZ$.

Most importantly, Newtonian physics seem to fail at a certain radial breakpoint, \Rd, where the acceleration is approximately \aZ.
Regarding this breakpoint as an acceleration threshold $a<\aZ$ leads to a search for a new physical laws which apply at low accelerations. Considering the equivalent threshold, $R>\Rd$, suggests that we don't understand what governs the behavior of the diffuse mass beyond the stellar disk and directs us to reconsider that region.

\section{Comparison to Previous Work} \label{sect:Cou07Comp}

The Table \ref{tbl:TheData} sample used here comprises only 26 galaxies and so it's appropriate to compare the current findings with the \cite{Cou07} study. \cite{Cou07} is a comprehensive study of the scaling relations of I-band luminosity, disk size, and rotational velocity in a sample of 1303 field and cluster disk galaxies and investigates the effect of color and morphological type and also includes a comparison to K-band results for a sub-sample of brighter galaxies.
\subsection{Description of the Data}
 \cite{Cou07} used I-band luminosity  from
 \begin{enumerate}
   \item The MAT sample of field galaxies\citep{Mat92},
   \item The SCII sample of cluster galaxies \citep{Dal99},
   \item The Shellflow study \citep{Cou00}, and
   \item The UMa survey citep{Tul96} \citep{Ver01}.
 \end{enumerate}

I-band disk scale length was used as a measure of disk size size and was measured somewhat differently for each sample.  Most of the measurements determine the scale length in the outer disk well away from any bulge or bar although ``erratic fits" were reported for some of the UMa measurements. Disc scale lengths were corrected for inclination using the formula
\begin{equation} R_I = \frac{R_{I,\obs}}{[1+0.4\log(a/b)]}.\end{equation}

Circular velocity was measured from resolved \Ha\ rotation curves, corrected for inclination and redshift broadening and was evaluated at 2.2 scale lengths except that for the SCII sample(about 40\% of the total) the circular velocity was evaluated at the optical radius.

I-band absolute magnitudes were calculated using $M_{I,\sun} = 4.19$ and were corrected for both internal and external extinction and a small k-correction was applied.  Distance was determined from the Hubble flow with H0=70h.

In order to perform a meaningful comparison it was necessary to convert the \cite{Cou07} data to the units used in the present work. The I-band luminosity and the radial scale length were multiplied by constant factors to convert to units of stellar mass and radial size.  The multipliers were chosen so that the centroid of the data lays on the line of intersection given by Equation \ref{equ:linedef}.  This procedure resulted in multipliers which are quite close to expected values:
\begin{itemize}
\item The I-band luminosity was converted to disk mass by multiplying by the factor \MLI =1.32.  This is close to the value of \MLI =1.2$\pm$0.2 given in \cite{Fly06}, especially considering that the \cite{Fly06} multiplier does not include an allowance for gas.
\item Scale length was converted to disk radius by multiplying by a factor of 4.05.  The conversion of scale length to radius is very uncertain, especially for this heterogeneous sample.  A value of about 3.2 is appropriate for an infinite exponential disk and larger value is expected for early type galaxies but depends on the details of the measurement.
\end{itemize}
Renormalizing the luminosity and scale factor data with these constant multipliers offsets the data in log space but doesn't affect the slope or scatter of the scaling relations.
\subsection{Evaluation}
Figure \ref{fig:Cou07}a-d  overlays Figures \ref{fig:M-V}-\ref{fig:M-RV} above onto the \cite{Cou07} data. As before, the solid lines show the Table  \ref{tbl:fundrel} relations with integer slopes and the dashed lines are $\pm 1 \sigma$ limits. Figures  \ref{fig:Cou07}a-c correspond to Figure 3 of \cite{Cou07} where I have suppressed the fitting lines and type information and transposed the axes of the M-V and M-R plots. Figure \ref{fig:Cou07}-d shows the MVR relation which was not discussed in the \cite{Cou07} analysis. The agreement is good showing a successful prediction of the Table \ref{tbl:fundrel} model.

Table \ref{tbl:CouStats} gives fitting statistics for each of the four relations.  The Pearson r coefficient indicates that the M-V and M-RV data can be fit quite well with a linear relation in log space.  The r coefficient for the M-R relation(0.70) and especially for the V-R(0.58) relation indicate that the fits are expected to be poor, which is the case.

Table \ref{tbl:CouStats} gives the scatter \sigln\ of each of the relations using the integer slopes given in Table \ref{tbl:fundrel} and also for optimized slopes which minimize \sigln.  The optimized slopes do reduce the scatter slightly but at the cost of introducing a new free variable.  The optimized slope of the M-RV relation is the coefficient $\gamma$ of the relation $M\varpropto\bigl(V^2 R\bigr)^\gamma$ which minimizes \sigln.  The reductions in \sigln\ which result from introducing additional free variables are small and parsimony rejects the optimized coefficients.

Equation \ref{equ:dispersion} can be used to approximate values of the random uncertainties of \Mt, \Vd\ and \Rd.  The implied dispersions are:

$\begin{array}{l  }
     0.10 < \textrm{u(\Mt)}  < 0.15  \\  
     0.02 < \textrm{u(\Vd)}  < 0.03  \\  
     0.15 < \textrm{u(\Rd)}  < 0.20  \\  
\end{array} $

These uncertainties are smaller than for previous large surveys but are significantly larger that the scatter of the Table \ref{tbl:TheData} data.  As expected, the scatter of \Rd\ is the the largest while the implied scatter of the \Ha\ rotational velocity is comparable to that of the \cite{McGau15} \HI\ data without the need to probe the outer gaseous disk.

The \cite{Cou07} data shown in Figure \ref{fig:Cou07} does show a deviation from the curve $M \propto V^4$ in that early-type galaxies fall on a line with a shallower slope.  The shallower slope might be due to the way that the circular velocity was measured.  That is,  \cite{Cou07} measured rotational velocity at an interior point of the disk(2.2 scale lengths) and the velocity is \emph{higher} there than at the edge for early-type galaxies but is \emph{lower} than at the edge for late-type galaxies. The effect might vanish if rotational velocity was measured consistently at the edge of the disk.
\begin{deluxetable}{lllll}[b!]
\tablecolumns{5}
\tablewidth{0pc}
\tablecaption{Fitting Statistics for \cite{Cou07} Data}
\tablehead{ \colhead{Description}&\colhead{M-V}&\colhead{M-R}&\colhead{R-V}&\colhead{M-RV}} \\
\startdata
\label{tbl:Cou07Stats}
Pearson r Coefficient     & 0.91     & 0.70     & 0.58    & 0.92   \\
\multicolumn{4}{c}{ ~ }     \\
Integer Slope             & 4        & 2        & 2       & 1      \\
Integer Slope \sigln      & 0.139    & 0.351    & 0.243   & 0.296  \\
 \multicolumn{4}{c}{ ~ }     \\
Optimized Slope           & 3.49     & 3.12     & 1.67    & 1.08   \\
Optimized Slope \sigln    & 0.133    & 0.310    & 0.241   & 0.291
\enddata
   \tablenotetext{ }{The optimized slopes minimize the rms scatter \sigln but result in only marginal improvement compared to the integer coefficients given in Table \ref{tbl:fundrel}.  }
\label{tbl:CouStats}
\end{deluxetable}
\section{Summary and Conclusions}\label{sect:Summary}
Two constants describe dynamics of stellar disks:
\begin{itemize}
\item The \Cf\ constant is the enhancement factor for a flat disk vs a spherical mass distribution. The value of \Cf\ derived here from observations is significantly larger than what is commonly applied and is close to the theoretical value. The database consists of mostly late-type galaxies and no morphological dependance of \Cf\ was found although other surveys, and standard Newtonian physics, indicate that the \Cf\ of  early-type galaxies is significantly smaller.
\item The \Aedg\ constant is the gravitational attraction at the edge of a stellar disk. The value of \Aedg\ is close to the value of Milgrom's constant but is not identical. Several possible explanations of why this attraction is constant have been proposed in the literature but the question is still open.
\end{itemize}

The  size of the stellar disk \Rd\ as defined here tied in a basic way to the dynamics of the disk. The small dispersion about the four observed correlations demonstrates that the fundamental relations of Equation \ref{equ:govern} are valid and that the procedure used to measure \Rd\ is meaningful.

The velocity curve beyond the stellar disk cannot be explained by standard Newtonian physics. Two alternative explanations are the consensus \LCDM\   theory and Milgrom's hypothesis. Although the current work does not disprove either explanation it casts some doubt on both. In particular, neither explanation assigns any particular significance to the size of the stellar disk which is, in fact, fundamentally important.
\section*{Acknowledgements}
This research has made use of the NASA/IPAC Extragalactic Database (NED) which is operated by the Jet Propulsion Laboratory, California Institute of Technology, under contract with the National Aeronautics and Space Administration. 

This work is based in part on observations made with the Spitzer Space Telescope, which is operated by the Jet Propulsion Laboratory, California Institute of Technology under a contract with NASA.

This work made use of the data sets used in \cite{McGau15} and \cite{Cou07} which the authors have made available to the community and is gratefully acknowledged.

The author would like to thank the anonymous referee for his valuable comments which helped to improve the manuscript. 

%

\end{document}